\documentclass[fleqn,usenatbib]{mnras}

\usepackage{newtxtext,newtxmath}

\usepackage[T1]{fontenc}

\usepackage{graphicx}	% Including figure files

\usepackage{amsmath}	% Advanced maths commands
\usepackage{amssymb}	% Extra maths symbols
\usepackage{siunitx}    % \num command for scientific notation
\usepackage{bm}         % bold math symbols
\usepackage[flushleft]{threeparttable}  % for table with notes
\usepackage{hhline}   % table formatting
\usepackage{orcidlink}

\title[TI and Multiphase Gas in the Simulated ISM]{Thermal Instability and Multiphase Gas in the Simulated Interstellar Medium with Conduction, Viscosity and Magnetic Fields}

\author[R. M. Jennings et al.]{R. Michael Jennings$^{1}$\thanks{E-mail: robertmjenningsjr@berkeley.edu}\orcidlink{0000-0002-3959-6572},
Yuan Li$^{1,2}$\orcidlink{0000-0001-5262-6150}
\\
$^{1}$Department of Astronomy, University of California, Berkeley, Berkeley, CA 94720\\
$^{2}$Department of Physics, University of North Texas, Denton, TX 76201\\
}

\date{Accepted XXX. Received YYY; in original form ZZZ}

\pubyear{\the\year}

\begin{document}
\label{firstpage}
\pagerange{\pageref{firstpage}--\pageref{lastpage}}
\maketitle

\begin{abstract}
Thermal instability (TI) plays a crucial role in the formation of multiphase structures and their dynamics in the Interstellar Medium (ISM) and is a leading theory for cold cloud creation in various astrophysical environments. In this paper we use two-dimensional (2D) simulations to investigate thermal instability under the influence of various initial conditions and physical processes. We experiment with Gaussian random field (GRF) density perturbations of different initial power spectra. We also enroll thermal conduction and physical viscosity in isotropic hydrodynamic and anisotropic magnetohydrodynamic (MHD) simulations. We find that the initial GRF spectral index $\alpha$ has a dramatic impact on the pure hydrodynamic development of thermal instability, influencing the size, number and motions of clouds. Cloud fragmentation happens due to two mechanisms: tearing and contraction rebound. In the runs with isotropic conduction and viscosity, the structures and dynamics of the clouds are dominated by evaporation and condensation flows in the non-linear regime, and the flow speed is regulated by viscosity. Cloud disruptions happen as a result of the Darrieus--Landau instability (DLI). Although at very late times, all individual clouds merge into one cold structure in all hydrodynamic runs. In the MHD case, the cloud structure is determined by both the initial perturbations and the initial magnetic field strength. In high $\beta$ runs, anisotropic conduction causes dense filaments to align with the local magnetic fields and the field direction can become reoriented. Strong magnetic fields suppress cross-field contraction and cold filaments can form along or perpendicular to the initial fields. 
\end{abstract}

\begin{keywords}
ISM: clouds -- instabilities -- plasmas -- MHD
\end{keywords}

%%%%%%%%%%%%%%%%%%%%%%%%%%%%%%%%%%%%%%%%%%%%%%%%%%

%%%%%%%%%%%%%%%%% BODY OF PAPER %%%%%%%%%%%%%%%%%%

\section{Introduction}

Thermal instability has been popularly employed to explain the existence of multiphase gas in a wide range of astrophysical contexts. The theory of TI was first seeded by \citet{Parker1953} by investigating the stability of equilibrium solutions to the heat equation, although proposed direct applications at the time did not venture beyond the condensations of solar prominences within the Sun's corona. Further investigations into the astrophysical outcomes of TI were done by \citet{Zanstra1955} where he suggested that the cool condensations seen in planetary nebulae exist in pressure equilibrium with the surrounding hotter gas and are formed through radiative cooling from forbidden line emission. In a milestone paper by \citet{Field1965} the linear theory of TI for a uniform gas in thermal equilibrium was first presented along with extensions discussing the effect of rotation, stratification, and magnetic fields. Later, \citet{Balbus1986} extended the model for a general case of a time dependant background flow.

The ISM is best described by a thermally bistable phase model consisting of two media of different densities and temperatures coexisting in pressure equilibrium. As was summarized by \citet{Spitzer1968}, ISM observations show \ion{H}{i} regions consisting of cool, dense clouds in pressure equilibrium with hot, rareified intercloud medium. \citet{Field1969} put forth the first description of the bithermal ISM phase model. In this model, TI is fueled by radiative cooling and cosmic ray heating to produce two thermally stable equilibrium phases--the cold neutral medium (CNM) and warm neutral medium (WNM)--separated by a thermally unstable phase. Later, the two phase model was extended with the inclusion of a hotter third phase heated by supernova explosions \citep{Cox1974,McKee1977,Cox2005} leading to the current multiphase theory of the ISM.  

For half a century now, TI has been the leading explanation for the existence of multiphase gas in diverse astrophysical contexts. From hundreds of kiloparsec scales within the intracluster medium and the circumgalactic medium (CGM) \citep{McCourt2012, Sharma2012, Li2014}, to tens of parsecs scales in the ISM, down to fractions of a parsec scales around accreting black holes, the ubiquity of a multiphase existence for astrophysical gas has been greatly appreciated. In the context of the ISM, TI has been an important ingredient when studying star formation as a sustaining creation process for the cold, dense molecular cloud star birthing regions \citep{Kim2013a,Kim2015}. Many early numerical works around TI centered on the question of whether it could sustain observed turbulence in the ISM \citep{VazquezSemadeni2000,Koyama2001,Kritsuk2002,Kritsuk2004}. Although consensus leans toward stellar feedback generated turbulence being dominant, it has been shown that TI turbulence alone can be completely self-sustaining \citep{Iwasaki2014}. 

A long sought after goal of TI studies has revolved around defining a characteristic size of clouds. An obvious minimum condensation size resulting from TI is the Field length, below which any density inhomogeneity is wiped out from thermal conduction. However, pinning down a characterstic cloud size has proven challenging. \citet{McCourt2018}  proposed and illustrated through 2D TI simulations a ``shattering'' mechanism where isochoric clouds rapidly fragment into cloudlets comparable to $\ell_{\text{cloudlet}} \sim c_st_{\text{cool}}$ to restore pressure equilibrium and continue cooling isobarically.  \citet{Waters2019a} reported ``splattering'' in TI due to acoustic oscillation. These shattering and splattering processes have been used to explain the host of current quasar line of sight absorption observations suggesting that cool galaxy halo gas comes in tiny, volume filling dense clouds \citep{McCourt2018}. In the ISM, suprathermal line widths of molecular clouds were previously thought to indicate internal velocities above the local sound speed. However, recent observations suggest that these broadened lines are composite subthermal lines of overlapping smaller cloudlets in relative motion to one another \citep{Tachihara2012}.

\citet{Gronke2020} quantitatively investigated the criteria for cloud shattering through 3D simulations arriving at the conclusion that small $\ell_{\text{cloudlet}}$ sized structures are seen after contraction rebound for clouds sufficiently out of pressure equilibrium $(\delta P / P \sim 1)$ and with final overdensities $\chi_f \equiv (\rho_{\text{CNM}}/\rho_{\text{WNM}})_f \gtrsim 300$. They also point out that the TI simulations of \citet{McCourt2018} that show shattering were initialized with highly non-linear conditions and thus show the first signs of shattering from linearily initialized TI. For clouds with smaller final overdensities they suggest that cloud coagulation is strong enough to stymie shattering and induce pulsations, bridging the gap between \citet{McCourt2018} and \citet{Waters2019a} explanations. \citet{Das2020} emphasize however, that the small cloudlet structures found by \citet{Gronke2020} are most likely formed as a result of Richtmyer--Meshkov instabilities \citep{Richtmyer1960,Meshkov1969} when shocks generate strong vorticity after rebound. They suggest that their own 1D TI simulations, along with the 3D TI simulations of \citet{Gronke2020}, do not follow the shattering process outlined by \citet{McCourt2018}. Despite lack of consensus on the validity of the shattering model, all recent TI simulations demonstrate cloud coagulation and the inevitable merging of cloudlets overtime. Therefore, regardless of how cloudlets form, great importance should be placed on the survivability and lifespan of small clouds. The cloud merging process exclusively has been studied in 2D simulations by \citet{Waters2019} and survivability has been pondered and investigated \citep{McKee1977,Nagashima2005,Nagashima2006,Nagashima2007,Yatou2009}.

Although \cite{Field1969} studied how magnetic fields impact the linear phase of TI more than half a century ago, the majority of numerical studies have neglected the magnetized case. The ubiquity of magnetic fields throughout astrophysical environments makes the consideration of magnetization essential, but the effects of magnetic fields can be complicated. Diffusion processes can differ greatly in the presence of magnetic fields due to the gyromotion of particles around the field lines. If the gyroradius, $r_g$, of a particle is much shorter than its collisional mean-free-path, $\lambda_{\text{mfp}}$, then diffusion is expected to become highly anisotropic. With heat transport being carried by electrons and momentum transport being carried by ions it is necessary to estimate these values for each particle species. For typical ISM values for the WMN, $\lambda_{\text{mfp}} \sim 10^{10}$cm is much larger than $r_{g,e} \sim 10^{6}$cm for electrons and $r_{g,i} \sim 10^{7}$cm for ions and so diffusion should be anisotropic. More recent numerical simulations have been carried out on anisotropic transport processes (conduction and viscosity) in the presence of magnetic fields \citep{Sharma2010,Choi2012a, Hennebelle2019}, but a large area of the parameter space remains to be explored. With observations showing possible correlation between cold filament orientation and magnetic field direction, TI involving these anisotropic transport processes necessitates numerical investigation. Given the existence of exquisite observations of the ISM, it is desirable to design and perform a systematic study on TI in the ISM to explore in detail the roles of different physical processes. This will also help inform our studies of TI in other astrophysical environments. 

In this paper we investigate TI under the influence of various initial conditions and physical processes. In isolating each process and its effect on TI evolution, we systematically explore the parameter space that TI operates within. We perform two-dimensional numerical hydrodynamic and MHD simulations of TI with physical conditions relevant for the ISM. In the absence of the microphysical processes of thermal conduction and physical viscosity we explore the importance of initial perturbations on the resulting structures and dynamics. For our choice of initial perturbations we emphasize the use of GRF density perturbations to maintain control over the power spectrum of wavelengths seeding the instability. We then investigate the addition of thermal conduction and physical viscosity and their effects. In the isotropic hydrodynamic cases we focus on the formation and the evolution of the multiphase structures. More specifically, we examine the merging and the breaking up of the cold clouds. In the MHD case we enroll anisotropic thermal conduction and viscosity. We explore the effects of varying magnetic field strengths in combination with varying initial perturbations. We study how resultant dense filaments are oriented relative to the fields.

This paper is organized as follows. In Section \ref{sec:methods} we introduce our numerical model and present a review of the linear theory of TI, followed with a 1D test and a description of the setup of our 2D simulations. In Section \ref{sec:hydrodynamic simulations}, we show results of our hydrodynamic simulations, both with and without thermal conduction and physical viscosity. For both cases, we discuss the various mechanisms for cloud disruption and cloudlet formation. We present MHD simulations with anisotropic thermal conduction and physical viscosity in Section \ref{sec:mhd}. We explore the effects magnetic fields have on the linear phase of TI, expanding on the analysis in \citet{Field1965}, and discuss the dependence of cloud structure on the field strength and initial perturbations. We summarize our work in Section \ref{sec:conclusion}.

\section{Methods} \label{sec:methods}
Our numerical models are given in Section \ref{sec:numerical model} along with a brief review of the multiphase model. The linear theory of TI is set up to test our model with 1D runs in Section \ref{sec:1d tests}. Our 2D simulation initial conditions are introduced in Section \ref{sec:2d simulations}.

\subsection{Numerical Model} \label{sec:numerical model}
We use the higher-order Godunov MHD code {\tt Athena++}\footnote{\href{https://github.com/PrincetonUniversity/athena-public-version/wiki}{github.com/PrincetonUniversity/athena-public-version/wiki}} \citep{Stone2020} with the HLLC (hydrodynamic runs) and HLLD (MHD runs) Riemann solvers, piecewise-parabolic reconstruction, and the 3rd order Runge--Kutta time integration algorithm. We solve the ideal MHD equations with the addition of thermal conduction, viscous transport, heating and cooling. The equations in conservative form are

\begin{equation} \label{mass}
    \frac{\partial \rho}{\partial t} + \bm{\nabla}  \cdot (\rho \bm{v}) = 0\,,
\end{equation}
\begin{equation} \label{mtm}
    \frac{\partial (\rho \bm{v})}{\partial t} + \bm{\nabla} \cdot \left[\rho \bm{vv} + \left(P + \frac{B^2}{8\pi}\right) \bm{I}  - \bm{\Pi}\right] - \frac{\bm{(B\cdot \nabla})B}{4\pi} = 0\,,
\end{equation}
\begin{equation} \label{energy}
    \frac{\partial E}{\partial t} + \bm{\nabla} \cdot \left[\bm{v}\left(E + P + \frac{B^2}{8\pi} \right) - \frac{\bm{B}(\bm{B} \cdot \bm{v})}{4\pi} + \bm{Q} - \bm{v} \cdot \bm{\Pi}\right] = - \rho\mathcal{L}\,,
\end{equation}
\begin{equation} \label{induction}
    \frac{\partial \bm{B}}{\partial t} - \bm{\nabla} \times (\bm{v} \times \bm{B}) = 0\,,
\end{equation}
which are the equations of mass, momentum and energy conservation and the induction equation without magnetic dissipation. The total energy $E$ is
\begin{equation}
    E = \mathcal{E} + \frac{1}{2}\rho v^2 + \frac{B^2}{8\pi}\,,
\end{equation}
where $\mathcal{E} = P/(\gamma -1)$ is the internal energy and we assume throughout this paper the adiabatic index value $\gamma = 5/3$ which corresponds to an ideal monoatomic gas. The heat flux $\bm{Q}$ and viscous stress tensor $\bm{\Pi}$ are composed of isotropic and anisotropic parts such that
\begin{equation}
    \bm{Q} = \bm{Q}_{\text{iso}} + \bm{Q}_{\text{aniso}} = -\left(\kappa_{\text{iso}}\bm{I} + \kappa_{\text{aniso}} \bm{\hat{b}\hat{b}}\right)\, \cdot\,  \bm{\nabla}T \,,
\end{equation}
and
\begin{multline}
    \bm{\Pi} = \bm{\Pi}_{\text{iso}} + \bm{\Pi}_{\text{aniso}} = \mu_{\text{iso}}\left[2\bm{D} - \frac{2}{3}(\bm{\nabla} \cdot \bm{v})\bm{I} \right] \\ + 3\mu_{\text{aniso}} \left[P_{\bot} - P_{\parallel}\right] \left[\bm{\hat{b}\hat{b}}- \frac{1}{3}\bm{I} \right]\,,
\end{multline}
where the deformation tensor $D$ is
\begin{equation}
    D_{ij} = \frac{1}{2}\left(\frac{\partial v_i}{\partial x_j} + \frac{\partial v_j}{\partial x_i}\right)\,,
\end{equation}
the pressure anisotropy is
\begin{equation}
    P_{\bot} - P_{\parallel} = \left[\bm{\hat{b}\hat{b}}:\bm{\nabla v} - \frac{1}{3}\bm{\nabla} \cdot \bm{v} \right]
\end{equation}
and $\bm{I}$ is the identity tensor. $\kappa$ and $\mu$ are the conduction and dynamic viscosity coefficients, respectively. We implement anisotropic conduction in an asymmetrical scheme with the monotonized-central slope limiter following the procedures of \cite{Sharma2007}. For anisotropic viscosity we use the discretization algorithm outlined in \citet{Parrish2012}. In runs where both thermal conduction and viscosity are employed, we signify the ratio between the two coefficients with the Prandtl number
\begin{equation}
    \text{Pr} = \left(\frac{\gamma}{\gamma -1}\frac{k_B}{\Bar{m}} \right)\frac{\mu}{\kappa}\,,
\end{equation}
where $k_B$ is Boltzmann's constant and $\Bar{m}$ is the mean mass per particle. Througout the paper we adopt the value $\Bar{m} = 1.27\,m_p$ which represents $10$ per cent by number He. The assumption of ideal MHD in essence relies on highly ionized plasma where the ions are completely tied to the magnetic field in a single fluid state. Although the temperature regimes of the ISM investigated in this paper involve only mildly ionized gas, the neutrals remain tied to the field via collisions with the ions. In reality we should expect the neutrals to drift from the ions and thus the field through ambipolar diffusion \citep{Hennebelle2019} but accounting for this complexity is beyond the scope of this paper.

\begin{figure}
    \centering
    \includegraphics[width=\columnwidth]{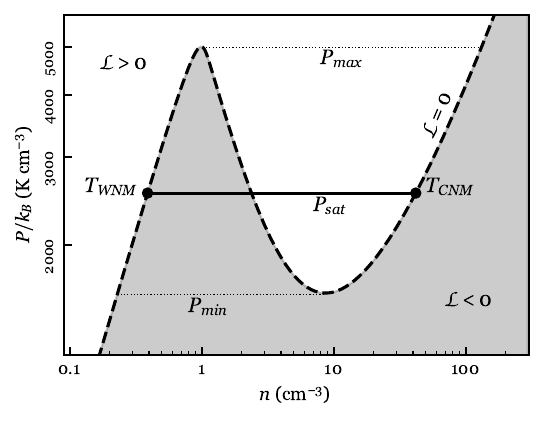}
    \caption{Equilibrium solutions $\mathcal{L} = 0$ for the chosen net cooling function. The $\mathcal{L} > 0$ portion above the equilibrium curve designates regions where cooling dominates heating, whereas the shaded $\mathcal{L} < 0$ portion designates regions where heating dominates cooling. The shape of the $\mathcal{L} = 0$ curve suggests that bistable configurations can only exist with pressure $P$ within the range $P_{\text{min}} < P < P_{\text{max}}$. When the environment pressure exceeds $P_{\text{max}}$ a cold stable phase can exist confined within an unstable gas. When the environment pressure is below $P_{\text{min}}$ condensations cannot form from a warm medium.}
    \label{fig:psat}
\end{figure}

A source term is added to the energy equation in the form of the net cooling function $\rho\mathcal{L} = n^2 \Lambda(T) - n\Gamma$ where $\Lambda$ and $\Gamma$ are the cooling and heating rates and $n = \rho/\Bar{m}$. We use the functional form suggested by \cite{Koyama2001},

\begin{multline}
    \frac{\Lambda(T)}{\Gamma} = 10^7 \exp{\left(\frac{\num{-1.184e5}}{T+1000}\right)} \\+ \num{1.4e-2}\sqrt{T}\exp{\left(\frac{-92}{T}\right)}\;\text{cm}^3\,,
\end{multline}
\begin{equation}
    \Gamma = \num{2e-26}\; \text{erg s}^{-1}\,.
\end{equation}
The heating term is comprised of photoelectric emission from small grains and polycyclic aromatic hydrocarbons \citep{Bakes1994}, ionization by cosmic rays and soft X-rays \citep{Wolfire1995}, and the formation and photodissociation of H$_2$ \citep{Hollenbach1979}. The cooling term is dominated by line emission from H, C, O, Si, and Fe, by rovibrational lines from H$_2$ and CO, and by atomic and molecular collisions with dust grains. The details of the heating and cooling contribution calculations are presented in \cite{Koyama2000}. 

Figure \ref{fig:psat} shows the equilibrium solutions $\mathcal{L} = 0$ with the given net cooling function. It can be seen by the shape of the $\mathcal{L} = 0$ curve that bistable configurations can only exist with pressure $P$ within the range $P_{\text{min}} < P < P_{\text{max}}$. When the environment pressure exceeds $P_{\text{max}}$ a cold stable phase can only exist confined within an unstable gas. This scenario is explored by \citet{Koyama2000} where a shock-compressed layer continually feeds the instability with thermally unstable gas. When the environment pressure is below $P_{\text{min}}$ condensations cannot form from a warm medium.

\subsection{1D Tests} \label{sec:1d tests}

In order to test the implementation of the heating, cooling
and conduction terms in our code, we have performed one-dimensional
simulations of the TI similar to the ones presented in \citet{Choi2012a,Piontek2004} and
compared the numerically measured growth rates with the theoretical prediction
first derived in \citet{Field1965}. 

The analytical growth rate of the TI comes from performing a linear wave analysis on the hydrodynamic equations with thermal conduction and cooling. By linearizing equations \eqref{mass}-\eqref{energy} minus the magnetic and viscous terms and assuming a small perturbation of the form $\,\propto\, \exp{[i(\bm{k} \cdot \bm{x} + \omega t)]}$ one can arrive at the dispersion relation for TI,
\begin{equation} \label{dispersion}
    \omega^3 + \omega^2c_s\left(k_T+\frac{k^2}{k_{\kappa}}\right)+\omega c_s^2k^2+ \frac{1}{\gamma}c_s^3k^2\left(k_T-k_{\rho}+\frac{k^2}{k_{\kappa}} \right) = 0\,,
\end{equation}
where the sound speed is given by $c_s = \sqrt{\gamma P/\rho}$. The characteristic wavenumbers are
\begin{equation}
    \begin{aligned}
        k_{\rho} &= (\gamma-1)\frac{\Bar{m}\rho \mathcal{L}_{\rho}}{c_sk_BT}\,,\\
        k_T &= (\gamma-1)\frac{\Bar{m} \mathcal{L}_{T}}{c_sk_B}\,,\\
        k_{\kappa} &= \frac{1}{(\gamma-1)}\frac{c_sk_B\rho}{\Bar{m}\kappa}\,,\\
    \end{aligned}
\end{equation}
where we have adopted the convenient notation $\mathcal{L}_\xi = (\partial \mathcal{L}/\partial \xi)$. As \citet{Field1965} points out, $\lambda_{\kappa} = 2\pi/k_{\kappa}$ represents the mean free path of the gas particles. For the cubic dispersion relation there are three solutions with one always being real; this real root signifies the exponential growth of what is called the \textit{condensation} mode and so the growth rate of the TI is given by the inverse of this root. The other two roots correspond to sound waves that are damped for typical cooling curves.

\begin{figure}
    \centering
    \includegraphics[width=\columnwidth]{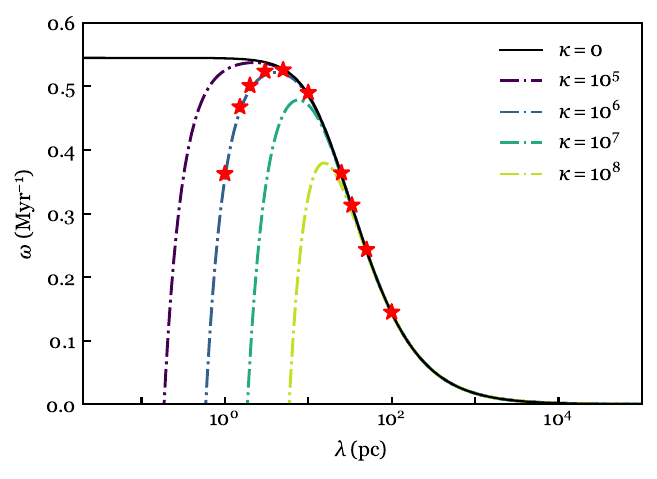}
    \caption{Growth rates $\omega$ of the one dimensional TI as a function of perturbation wavelength $\lambda$ for different values of constant thermal conduction coefficient $\kappa$. The dash-dotted lines representing TI in the presence of thermal conduction cross zero at the Field length $\lambda_F$. The red stars show the numerically measured growth rate of our test simulations with $\kappa = 10^6$.}
    \label{fig:1D_Test_Results}
\end{figure}

The initial conditions for this test are a medium
at rest with constant density and pressure of $n = 2.0\, \text{cm}^{-3}$ and
$P/k = 3000\, \text{K cm}^{-3}$, and isotropic conduction with coefficient
$\kappa = 10^6\,\text{erg s}^{-1}\,\text{cm}^{-1}\,\text{K}^{-1}$. The computational domain size is $L = 100\, \text{pc}$ and
the grid contains 2048 zones with periodic boundaries. We initialize the models with
eigenmodes of the instability by imposing sinusoidal fluctuations
of amplitude $A = 0.1$ per cent and with wavenumber $k$. The following equations outlined in \cite{Waters2019a} serve as our initial conditions for the 1D test simulations:
\begin{equation}
    \begin{aligned}
        \rho(\bm{x},t) &= \rho_0 + A\rho_0\cos{(\bm{k\cdot x})}\,, \\
        v(\bm{x},t) &= -A\left(\frac{\omega}{k}\right)\sin{(\bm{k\cdot x})}\,, \\
        P(\bm{x},t) &= P_0 - A\rho_0\left(\frac{\omega}{k}\right)^2\cos{(\bm{k\cdot x})}\,.
    \end{aligned}
\end{equation}

In order to analyze the linear regime of the growth the simulation was run for $10^{-2}\,t_{\text{cool},0}$, where $t_{\text{cool}}$ is defined as the time it takes for a gas to lose its internal energy due to cooling,
\begin{equation}
    t_{\text{cool}} \equiv \frac{\mathcal{E}}{n\Lambda}\,.
\end{equation}
For our initial conditions $t_{\text{cool},0} \approx 0.48 \text{ Myr}$. This cooling time implies a characteristic length scale termed the cooling length,
\begin{equation}
    \lambda_{\text{cool}} \equiv c_s t_{\text{cool}}\,,
\end{equation}
which for our initial conditions gives a value $\lambda_{\text{cool},0} \approx 2$ pc. The numerical growth rate $\omega_{\text{num}}$ was then obtained by measuring the logarithmic rate of change of the maximum density,
\begin{equation}
    \omega_{\text{num}} = \frac{d}{dt}\log{\left(\frac{\rho_{\text{max}}}{\rho_0}-1\right)}\,.
\end{equation}

Figure \ref{fig:1D_Test_Results} compares the numerical growth rates from
the 1D test simulations to the theoretical values for the coefficient of thermal conductivity used for all of the runs throughout this paper. Also shown are lines for other values of $\kappa$ to illustrate how characteristic values of the TI would change. Most notably, without thermal conduction, the fastest growing modes are on the smallest scales. The wavelength at which growth is completely suppressed by conduction is the Field length,
\begin{equation}
    \lambda_F \equiv 2\pi \sqrt{\frac{\kappa T}{n^2 \mathcal{L}_M}}\,,
\end{equation}
where $\mathcal{L}_M \equiv \text{max}(\Lambda,\Gamma/n)$ \citep{Begelman1990}. The wavelength where growth is maximal can be estimated as the geometric mean of the two characteristic lengths,
\begin{equation}
    \lambda_{\text{max}} \approx \sqrt{\lambda_{\text{cool}} \lambda_F}\,.
\end{equation}

\subsection{2D Simulations} \label{sec:2d simulations}
We seed the TI with multiple isobaric modes by generating a GRF of the following form:
\begin{equation} \label{grf}
    \delta (x,y) = \sum^{i_{\text{max}}}_{i,j=0} (k_i^2+k_j^2)^{\frac{\alpha}{2}}\sin{[xk_i+\phi(k_i,k_j)]}\sin{[yk_j+\phi(k_i,k_j)]}\,,
\end{equation}
where $k_i = i \times 2\pi / L$ is the wavenumber, $\alpha$ is the spectral index and $\phi$ is a random phase. For all runs we take $i_{\text{max}} = 40$. We normalize the GRF to have a root mean square (RMS) value equal to the desired amplitude $A$,
\begin{equation} \label{RMS}
    \text{RMS} = \sqrt{\frac{1}{L_xL_y}\sum^{L_x,L_y}_{x,y=0} \delta (x,y)^2} = A\,.
\end{equation} 

\begin{figure}
    \centering
    \includegraphics[width=\columnwidth]{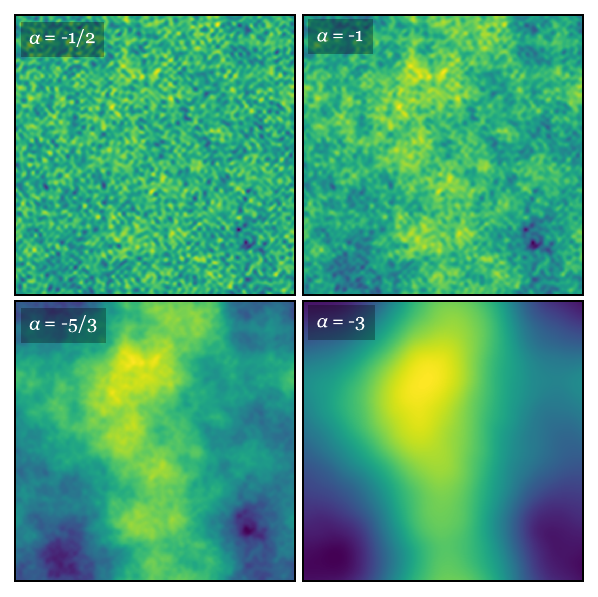}
    \caption{Initial Gaussian random field (GRF) isobaric density perturbations with spectral indices $\alpha = -1/2,-1,-5/3,-3$ at $t=0$. All simulations presented in this paper use one of these initial perturbations unless otherwise stated.}
    \label{fig:spec_ind}
\end{figure}

The density inhomogeneities in the ISM have been shown to follow a Kolmogorov spectrum \citep{Armstrong1994,Chepurnov2010}, so we take a spectral index of $\alpha = -5/3$ for our fiducial runs to best mimic the turbulent conditions that would instigate TI collapse. Initial density perturbations for different spectral indices are shown in Figure \ref{fig:spec_ind}. All simulations presented in this paper start with initial density perturbations matching one of the four shown depending on the chosen spectral index unless otherwise stated. We also run several simulations seeded by cell-to-cell random density perturbations that mimic a ``white noise'' approach for comparison. 

\begingroup
\setlength{\tabcolsep}{29pt} 
\renewcommand{\arraystretch}{1.1} 
\begin{table*}
\centering
\caption{2D Thermal Instability Models}
\begin{threeparttable}

\label{table:models}
\begin{tabular}{lccccc}
  \hhline{======}
  Model\tnote{a} & $N$\tnote{b} & $L$\tnote{c} & Pr\tnote{d} & $\beta$\tnote{e} & $\alpha$\tnote{f} \\
  \hline
  Na05 & 800 & 8pc & - & - & $-1/2$ \\
  Na1 & 800 & 8pc & - & - & $-1$ \\
  Na5/3 & 800 & 8pc & - & - & $-5/3$ \\
  Na3 & 800 & 8pc & - & - & $-3$ \\
  \hline
  pr1e-2w & 2048 & 4pc & $10^{-2}$ & - & white noise \\
  pr1/2w & 2048 & 4pc & $1/2$ & - & white noise \\
  pr2/3w & 2048 & 4pc & $2/3$ & - & white noise \\
  pr8e-1w & 2048 & 4pc & $0.8$ & - & white noise \\
  pr9e-1w & 2048 & 4pc & $0.9$ & - & white noise \\
  pr1w & 2048 & 4pc & $1$ & - & white noise \\
  \hline
  pr1e-2a05 & 2048 & 4pc & $10^{-2}$ & - & $-1/2$ \\
  pr1e-2a1 & 2048 & 4pc & $10^{-2}$ & - & $-1$ \\
  pr1e-2a5/3 & 2048 & 4pc & $10^{-2}$ & - & $-5/3$ \\
  pr1e-2a3 & 2048 & 4pc & $10^{-2}$ & - & $-3$ \\
  pr2/3a5/3 & 2048 & 4pc & $2/3$ & - & $-1/2$ \\
  pr2/3a1 & 2048 & 4pc & $2/3$ & - & $-1$ \\
  pr2/3a5/3 & 2048 & 4pc & $2/3$ & - & $-5/3$ \\
  pr2/3a3 & 2048 & 4pc & $2/3$ & - & $-3$ \\
  \hline
  b1e6pr1e-2a05 & 2048 & 4pc & $10^{-2}$ & $10^{6}$ & $-1/2$ \\
  b1e6pr1e-2a1 & 2048 & 4pc & $10^{-2}$ & $10^{6}$ & $-1$ \\
  b1e6pr1e-2a5/3 & 2048 & 4pc & $10^{-2}$ & $10^{6}$ & $-5/3$ \\
  b1e4pr1e-2a05 & 2048 & 4pc & $10^{-2}$ & $10^{4}$ & $-1/2$ \\
  b1e4pr1e-2a1 & 2048 & 4pc & $10^{-2}$ & $10^{4}$ & $-1$ \\
  b1e4pr1e-2a5/3 & 2048 & 4pc & $10^{-2}$ & $10^{4}$ & $-5/3$ \\
  b1pr1e-2a05 & 2048 & 4pc & $10^{-2}$ & $1$ & $-1/2$ \\
  b1pr1e-2a1 & 2048 & 4pc & $10^{-2}$ & $1$ & $-1$ \\
  b1pr1e-2a5/3 & 2048 & 4pc & $10^{-2}$ & $1$ & $-5/3$ \\
  \hline
\end{tabular}
\begin{tablenotes}
  \item[a] Model naming convention: N = no conduction/viscosity, aXX = spectral index -XX, prXX = Prandtl number XX, bXX = initial plasma beta XX, w = initialized with random white noise density fluctuations instead of GRF.
  \item[b] Number of grid points $N = N_x = N_y$.
  \item[c] Physical length of box $L = L_x = L_y$.
  \item[d] Prandtl number with fixed $\kappa$ and variable $\mu$.
  \item[e] Initial plasma beta.
  \item[f] Spectral index for initial density Gaussian random field power spectrum. GRF density perturbations are not used for models labeled ``white noise'' and are instead initialized with cell-by-cell random density fluctuations.
\end{tablenotes}
\end{threeparttable}
\end{table*}
\endgroup

The initial number density and pressure are the same as our 1D test simulations. The domain for all our runs are square with $L_x = L_y = L$ with periodic boundary conditions. For all runs with thermal conduction and physical viscosity we take the constant value $\kappa = \num{9.68e4}\,\text{erg s}^{-1}\,\text{cm}^{-1}\,\text{K}^{-1}$ corresponding to a temperature of $T = 1500 \,\text{K}$ according to \citep{Parker1953, Spitzer1962} 
\begin{equation} \label{spitz cond}
    \kappa = \num{2.5e3}\, T^{1/2} \, \text{erg s}^{-1}\,\text{cm}^{-1}\,\text{K}^{-1}\;,
\end{equation}
and adjust $\mu$ for the desired value of the Prandtl number. We ensure that the shortest CNM Field length is resolved by at least 3 zones according to the condition suggested by \citet{Koyama2004}; for runs without thermal conduction we ensure at least the CNM cooling length is resolved. All runs that enroll thermal conduction also enroll viscosity (for both isotropic and anisotropic cases) for two main reasons: (1) to ensure numerical convergence of the density and velocity distributions, especially in the evaporation/condensation zones \citep{Choi2012a}; (2) to ensure proper comparison between runs in the saturation phase since the saturation velocity is set by the balance of evaporation due to thermal conduction and momentum diffusion due to viscosity \citep{Koyama2006}.

For MHD runs the magnetic field is initialized at an angle $\theta_0 = 45^{\circ}$ from the $x-$axis. In order to ensure numerical convergence and to limit the growth of certain plasma instabilities when employing anisotropic transport processes, an additional small isotropic coefficient is required. The smaller isotropic thermal conduction component introduces a Field length perpendicular to the magnetic field lines that must also be resolved to avoid resolution dependent filament thicknesses \citep{Sharma2010}. We adopt the value $\kappa_{\text{iso}} = \num{3.23e3}\,\text{erg s}^{-1}\,\text{cm}^{-1}\,\text{K}^{-1}$ which gives a ratio $\kappa_{\text{ani}}/\kappa_{\text{iso}} \approx 30$. This artificially large value for thermal conduction coefficient perpendicular to the field will cause a systematic overestimate of cross-field velocities as well as filament thickness, however reducing this value would make computations extremely restrictive. 

A magnetized plasma becomes subject to microscale instabilities when the pressure anisotropy exceeds the range \citep{Kunz2012}
\begin{equation} \label{pressure_ani}
    -\frac{B^2}{4\pi} \lesssim P_{\bot} - P_{\parallel} \lesssim \frac{B^2}{8\pi} \,.
\end{equation}
These instabilities tend to grow to correct for the excess pressure anisotropy that creates them. However, growth of these instabilities (e.g. firehose and mirror instabilities) occurs fastest for wavelengths at the grid scale which often means they are unresolved in simulations. Employing anisotropic viscosity introduces the danger of unregulated pressure anisotropies and so an artificial isotropic viscosity is necessary to limit the pressure anisotropies \citep{Schekochihin2005}. We introduce $\mu_{\text{iso}}$ such that the Prandtl number for the isotropic coefficients is the same as for the anisotropic coefficients.

\section{Hydrodynamic Simulations} \label{sec:hydrodynamic simulations}
\begin{figure*}
    \centering
    \includegraphics[width=\textwidth]{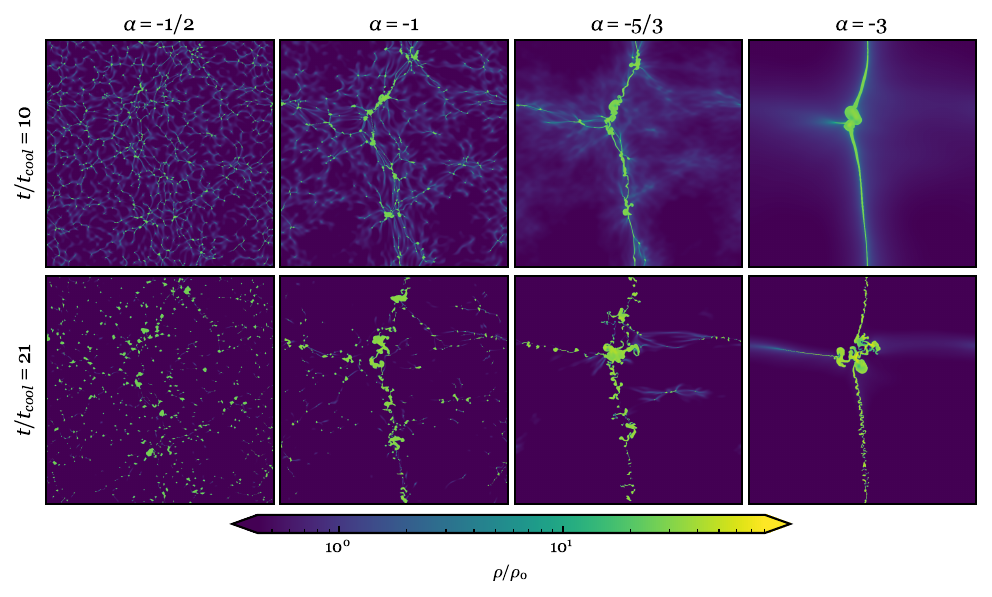}
    \caption{Pure hydrodynamic runs with GRF density perturbations of different spectral indices $\alpha$. For smaller $|\alpha|$ power is concentrated in smaller wavelength perturbations causing the formation of smaller and more numerous cloud structures. For larger $|\alpha|$ power is concentrated in larger wavelength perturbations, resulting in larger and fewer cloud structures. The top and bottom sets show snapshots at two different times. Cloud fragmentation is easily seen for the larger magnitude spectral indices at late times (bottom panels). Most notably,  filaments in the $\alpha = -3$ snapshot are torn apart, which we discuss further in Section~\ref{sec:nocond_frag} and Figure~\ref{fig:fragment}.}
    \label{fig:nocond1}
\end{figure*}

We present the results of simulations with various physics implemented in order to investigate their direct effects on cloud formation and evolution. In Section~\ref{sec:nocond} we first present results of pure hydrodynamic runs without conduction and viscosity. As mentioned previously, without thermal conduction TI grows most rapidly on the smallest scales and thus the evolution of the instability is dependant on the resolution as well as the modes present in the initial perturbation. We use these runs to investigate how large of an effect the initial perturbation has on the subsequent cloud formation. We then present the results of hydrodynamic runs with conduction and viscosity. All simulation models are listed in Table \ref{table:models}.

\subsection{Hydrodynamic Simulations without Conduction} \label{sec:nocond}
\begin{figure*}
    \centering
    \includegraphics[width=\textwidth]{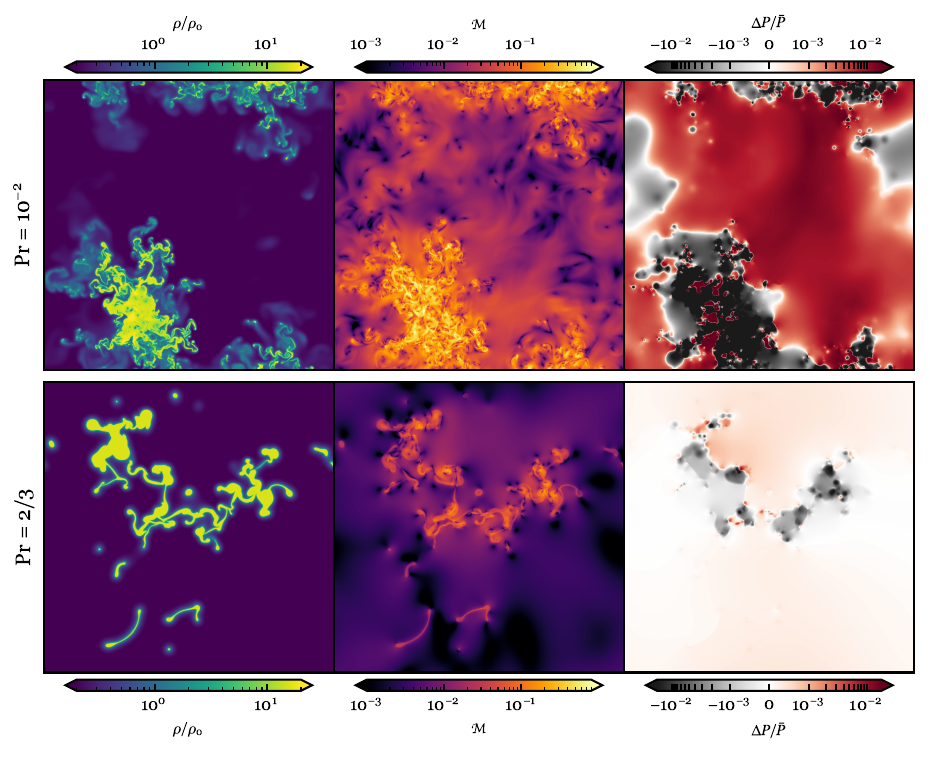}
    \caption{Hydrodynamic simulations with conduction and viscosity at late times of $t/t_{\text{cool}} \approx 280$ with $\text{Pr} = \num{1e-2}$ (top panels) and $2/3$ (bottom panels). The plots show, starting with the left panel: density contrast $\rho/\rho_0$, Mach number $\mathcal{M} = v/c_s$, and deviation from the mean pressure $\Delta P/\Bar{P}$. In the run with $\text{Pr} = \num{1e-2}$ the thermal conduction dominates over the viscous dissipation and aggressive evaporation from cloud fronts contributes to a highly turbulent gas in both the CNM and WNM. The evaporation causes large pressure gradients that drive cloud motion, resulting in rapid and continuous cloud merging. In the run with $\text{Pr} = 2/3$ viscous dissipation reduces the evaporation rate considerably, producing a less turbulent environment with less deviation from pressure equilibrium.}
    \label{fig:iso}
\end{figure*}

We first conduct pure hydrodynamic TI runs initialized with cell-by-cell random density perturbations, following previous works by \citet{Choi2012a} and \citet{Piontek2004}. By introducing the instability seed this way, the perturbation is dominated on the resolution scale. Therefore, the linear phase of TI produces cell size clouds that later merge in the non-linear phase. 

To investigate the effect of different initial perturbations, we run a set of pure hydrodynamic simulations with GRF perturbations of various initial spectral indices.
Density plots ($\rho/\rho_0$) at $t / t_{\text{cool}} = 10.4$ are shown in Figure \ref{fig:nocond1}. It is clear that the size and number of initial clouds formed from TI is highly dependant on the power spectrum used to seed the perturbations. A small spectral index $|\alpha|$ puts the majority of power in perturbations shorter than the cooling length. Therefore, the condensation mode is able to evolve isobarically allowing sound waves to restore pressure equilibrium. In this case, large pressure gradients are wiped out and cloud merging happens slowly due to lower velocities. In the extreme case where $|\alpha|$ tends toward zero, the results from the random white noise perturbations of \citet{Choi2012a} are reproduced. With spectral indices increasing in magnitude (more negative), power is pushed into perturbations larger than the cooling length. Condensation modes collapse isochorically, allowing for cooling to occur before sound waves can restore pressure equilibrium. In this case, larger pressure gradients result in higher velocities and quicker cloud merging. Figure \ref{fig:nocond1} shows that in the non-linear phase, small $|\alpha|$ results in smaller structures and greater numbers of clouds, while large $|\alpha|$ produces larger structures and fewer numbers of clouds.
Although at much later times ($ t/t_{\text{cool}} > 50$, not shown in Figure \ref{fig:nocond1}), all runs look similar due to cloud merging. The early evolution may be very important in the realistic conditions of the ISM where turbulence can disrupt clouds before substantial merging takes place.

\subsection{Hydrodynamic Simulations with Isotropic Conduction and Viscosity}

We run simulations with thermal conduction and viscosity with varying spectral indices, and find that conduction quickly dominates the evolution of the system. The results are almost independent of $|\alpha|$ except at the very beginning of the TI. Therefore, we concentrate our further analysis on runs with the fiducial spectral index $\alpha = -5/3$.

We present non-linear results at late times of $t/t_{\text{cool}} \approx 280$ with $\text{Pr} = 10^{-2},\;2/3$ in Figure \ref{fig:iso}, including the density contrast $\rho/\rho_0$, Mach number $\mathcal{M} = v/c_s$, and deviation from the mean pressure $\Delta P/\Bar{P}$. 
In the run with $\text{Pr} = 10^{-2}$ (top panels of Figure \ref{fig:iso}), the thermal conduction dominates the viscous dissipation. Aggressive evaporation from cloud fronts contributes to a highly turbulent gas in both the CNM and WNM. The evaporation causes large pressure gradients that drive cloud motion, resulting in rapid and continuous cloud merging. Quickly after TI saturation, the bulk of the cold gas has coalesced into one large cloud restricted in size by the box boundaries. As the Prandtl number is increased towards unity, viscous dissipation begins to rival the evaporation flow from conduction, and turbulence decreases in the $\text{Pr} = 2/3$ run (bottom panels of Figure \ref{fig:iso}). Consequently, the smaller pressure gradients reduce the velocity of the gas, slowing down cloud merging. This results in the longevity of various disconnected cold structures throughout the box. We find that increasing Prandtl number also reduces the fraction of mass lying within the unstable region, shrinking the width of the interfacial boundary layer defined by $(\partial \mathcal{L}/\partial{\rho})_P > 0$ \citep{Iwasaki2014}, consistent with \citet{Choi2012a}.

\begin{figure}
    \centering
    \includegraphics[width=\columnwidth]{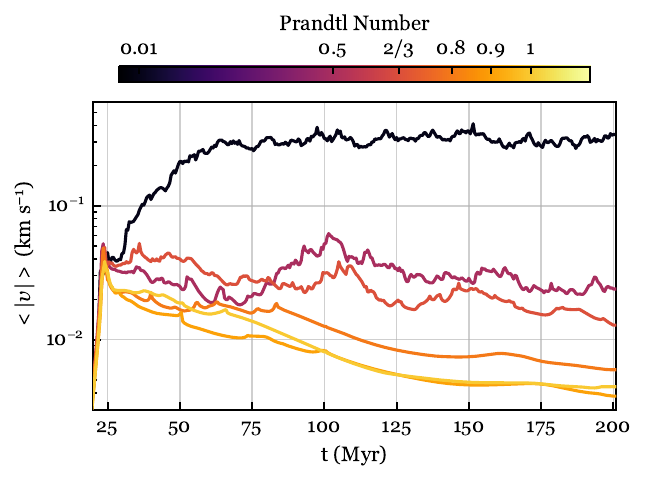}
    \caption{Mean mass-weighted velocity evolution in the hydrodynamic runs with various Prandtl numbers. Viscous dissipation becomes stronger with increasing Prandtl number. Oscillations in the mean velocity occur on finer timescales when thermal conduction is more dominant. In higher Prandtl number runs, those finer oscillation structures are effectively smoothed by viscous dissipation.}
    \label{fig:velevol}
\end{figure}

The mean velocity evolution of various Prandtl numbers are shown in Figure \ref{fig:velevol} with the white noise seeded initial conditions. We choose to compare these runs to ensure that the main velocity contributions are from evaporation flows and not from the collapse of any isochoric modes. We find that the saturation velocity of TI is a decreasing function of Prandtl number, in agreement with \citet{Koyama2006}. This also explains the slower cloud merging and better survival rates in high Prandtl number runs. In addition, viscosity reduces the velocity short timescale oscillations in the non-linear phase by suppressing the evaporation flows, resulting in smoother time evolution of the mean velocity with higher Prandtl numbers.

\subsection{Cloud Disruption in Hydrodynamic Simulations}

\subsubsection{Fragmentation without Thermal Conduction} \label{sec:nocond_frag}

Cloud fragmentation in pure hydrodynamic simulations has been the focus of discussion in many recent works \citep[e.g.,][]{McCourt2018, Gronke2020, Das2020}. Our pure hydrodynamic runs without thermal conduction also show clear signs of cloud fragmentation. We recognize two processes for clouds to break into smaller cloudlets: a Richtmyer--Meshkov like propulsion of material outward from the main cloud after contraction rebound, and fragmentation of smaller clouds by pressure gradients between two or more larger clouds during the merging process. Both processes shred the clouds apart via large vorticity, $\bm{\nabla} \times \bm{v}$, and are fundamentally distinct from the \citet{McCourt2018} ``shattering'' description.

The first mechanism of Richtmyer--Meshkov-like fragmentation was presented in \citet{Gronke2020} for clouds with final overdensities $\chi_f  \gtrsim 300$ where non-uniformly shaped dense protrusions are ejected from the larger cloud during contraction rebound. The ejection's interaction with the warm expanding medium causes strong vorticity that breaks the cloud into smaller cloudlets. This process is seen for all four spectral indices runs but is less pronounced for the $\alpha = -1/2$ run due to very small velocities from nearly in-place isobaric condensations. 

The second mechanism of clouds being shredded by opposing larger cloud masses during merging is discussed less in previous work. In the non-linear phase of TI, gas within the interfacial layer continues to cool and heat, creating pressure gradients that induce cloud motions. As \citet{Waters2019} show, two spherical clouds always merge in their 2D simulations. In systems containing  multiple clouds with various sizes and geometries, a small cold cloud caught within the coalescing influence of two large neighboring clouds can be torn apart. The competing pressure gradients of the two neighboring clouds generate strong vorticity that shreds the smaller cloud. This process occurs in all four of our spectral index runs. 

\begin{figure}
    \centering
    \includegraphics[width=\columnwidth]{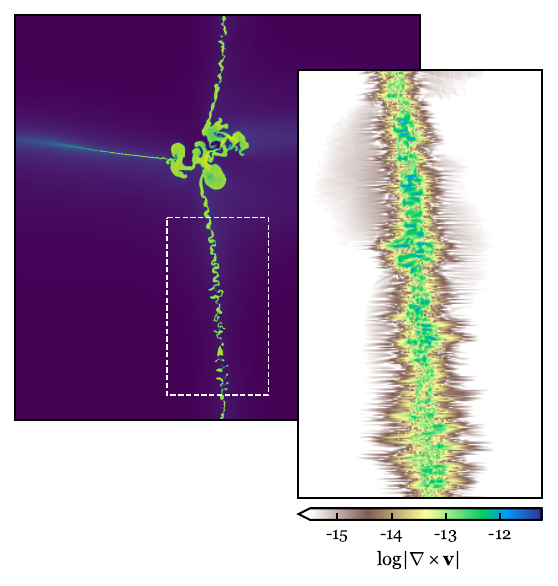}
    \caption{Fragmentation of a cold filament in the early nonlinear phase of the $\alpha = -3$ hydrodynamic run without conduction or viscosity. Inhomogeneities in the collapse process create high vorticity in an unstable cold filament that is more susceptible to fragmentation. The fragmentation begins as an oscillatory disturbance that propagates down the length of the filament eventually ripping it apart into smaller cloudlets. This behavior was seen in all $\alpha$ runs but was most apparent and occurred on a larger scale in the $\alpha = -3$ run due to larger pressure gradients at the interfacial layers.}
    \label{fig:fragment}
\end{figure}

An example where both mechanisms take place is shown in Figure \ref{fig:fragment}. Dense filaments form above and below the main central cloud condensation. With periodic boundary conditions the filament exists between two large dense clouds, both exerting a pressure gradient to pull the filament material into themselves. The filament is already unstable in the left-right directions from inhomogeneities in the condensation process which cause high vorticity. The filament then shreds itself apart from the increased internal vorticity and coalescent influence of the larger central cloud. A majority of the smaller cloudlets that form are pulled into the central cloud. The few cloudlets lying near the box boundaries remaining somewhat stagnate under symmetrical dominant pressure gradients due to the periodic conditions.

The cloud sizes depend heavily on the initial perturbations during the linear phase of TI. In the non-linear phase, we see a variety of cold structures with a wide range of sizes in all our pure hydro runs. Even at very late times when all cold gas merges into one large cloud, the cloud has features of much smaller sizes. In other words, we do not see a characteristic size as described in \citet{McCourt2018}, and definitely not at $c_st_{\text{cool}}$. Although we note that our simulation setup is quite different and so are the initial perturbations. 

\subsubsection{Fragmentation with Thermal Conduction: Geometric Disturbances and DLI}

In idealized 1D calculations, clouds smaller than $\lambda_F$ always evaporate due to conductive transport dominating heating/cooling rates \citep{McKee1990}. If the pressure is sufficiently higher than the saturation pressure $P_{\text{sat}}$, clouds larger than $\lambda_F$ will grow from the gas in the interfacial layer condensing. The saturation pressure is the pressure that allows for a stationary transition layer between the cold and hot phases where heat conduction compensates the heating/cooling imbalance throughout the unstable regime:
\begin{equation} \label{psat}
    \int_{T_{\text{CNM}}}^{T_{\text{WNM}}}\kappa \rho \mathcal{L} dT = 0\;,
\end{equation}
\citep{ZelDovich1969,Penston1970}. If the pressure is below that of $P_{\text{sat}}$ then evaporation occurs; for pressure above $P_{\text{sat}}$ condensation occurs. The density-pressure phase diagram in Figure \ref{fig:psat} illustrates these characteristic values with the given equilibrium curve where $\mathcal{L} = 0$. 

The surface of conductive clouds is subject to an instability analogous to the DLI studied in terrestrial flames. As \citet{Inoue2006, Kim2013} show, an initially infinite planar-parallel, stationary evaporation/condensation transition layer will experience DLI with any sufficiently large ($>\lambda_F$) wavelength distortion to the front. Consider a distorted front such that the cold cloud protrudes into the warm medium. In the case of an evaporating cloud, the cold gas that flows across the front will refract toward the normal due to expansion, increasing the local pressure and mass flux. This increased mass flux necessitates and advancement of the protrusion creating a cold finger. In contrast, the locations adjacent to the protrusion, or the cusps, will decrease in pressure due to the opposite process. Eventually, non-linear DLI saturates after long finger-like protrusions form.

\begin{figure}
    \centering
    \includegraphics[width=1.0\columnwidth]{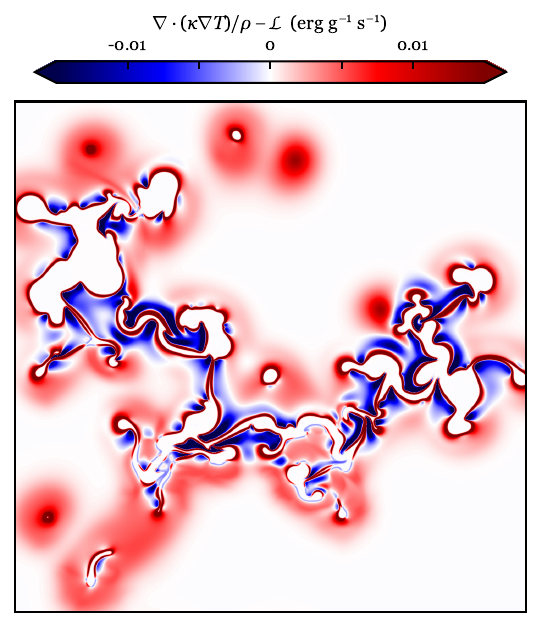}
    \caption{The sum of the radiative heating/cooling and thermal conduction terms for a zoomed-in portion ($2.8\,\text{pc}\, \times\, 2.8\,\text{pc}$, a 1.4$\times$ zoom factor) of the $\text{Pr} = 2/3$ isotropic run. The curvature $K$ of the cloud surface is highly influential on the conduction term with $K>0$ stimulating evaporation and $K<0$ stimulating condensation. Convex portions are dominated by gas heating which causes overpressured regions while concave portions are dominated by gas cooling which causes underpressured regions. The created pressure gradients in turn influence the gas motions.}
    \label{fig:enthalpy}
\end{figure}

In the simulations presented in this paper, full cloud geometry breaks the symmetry of an infinite planar idealization. The criteria for gas evaporating/condensation are determined by the energy exchange during the phase transition. The dominate thermal processes are radiative heating/cooling $-\mathcal{L}$ and thermal conduction $\nabla \cdot (\kappa \bm{\nabla T})/\rho$. \citet{Nagashima2005} showed that curved fronts experience stimulated evaporation/condensation based on the curvature $K = \bm{\nabla} \cdot \bm{\hat{n}}$, where $\bm{\hat{n}} = \bm{\nabla T}/|\nabla T|$. Convex (Concave) portions of cold clouds that have $K > 0\; (K<0)$ experience stimulated evaporation (condensation). This can be seen by decomposing the thermal conduction term into a parallel component and a curvature component,
\begin{equation} \label{curv}
    \bm{\nabla} \cdot (\kappa \bm{\nabla T}) = \partial_n(\kappa \partial_n T) + \kappa K \partial_n T\;,
\end{equation}
where $\partial_n \equiv \bm{\hat{n}} \cdot \bm{\nabla}$ \citep{Iwasaki2014}. Therefore, which direction energy is transferred within the phase transition front is highly dependant on the curvature of the cloud surface. Figure \ref{fig:enthalpy} shows the sum of the radiative heating/cooling and thermal conduction terms for a zoomed-in portion ($2.8\,\text{pc} \times 2.8\,\text{pc}$) of the $\text{Pr} = 2/3$ isotropic run. The convex portions are dominated by gas heating resulting in overpressured regions, whereas concave portions are dominated by gas cooling which produces underpressured regions. The resulting pressure gradients influence the gas motions. This is what determines evaporation/condensation for our simulations, in contrast to the $P/P_{\text{sat}}$ relationship presented in 1D steady solution analysis.

\begin{figure}
    \centering
    \includegraphics[width=\columnwidth]{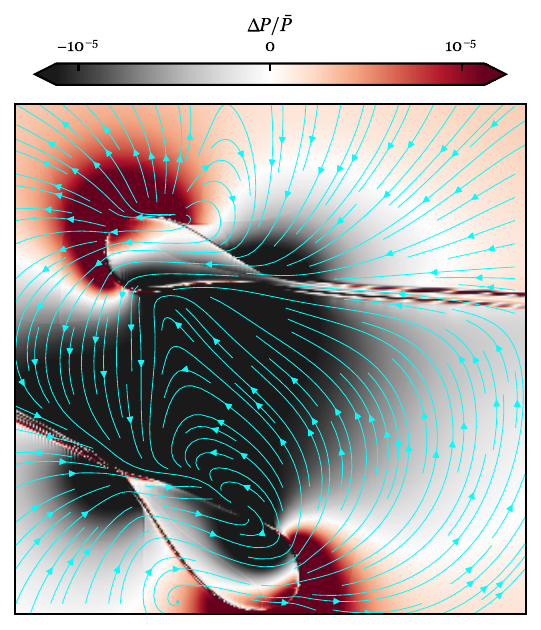}
    \caption{Deviation from the mean pressure $\Delta P/\Bar{P}$ for a zoomed-in portion ($1\,\text{pc}\, \times\, 1\,\text{pc}$, a 4$\times$ zoom factor) of a $\text{Pr} = 2/3$ run. Overlaid are velocity streamlines. 
    High pressure is seen around the convex regions of the bulbs, and is created from stimulated evaporation due to the $K > 0$ curvature effect on thermal conduction. Lower pressure is seen at the concave regions just behind the bulb where condensation is stimulated by $K < 0$ curvature effects. The configuration of these over and under pressured regions influences mass flow into the cusps behind the bulbs. This momentum flux causes the bulbs to break off to form independent cloudlets.}
    \label{fig:dli}
\end{figure}

The complex interplay between curvature-stimulated evaporation and condensation ultimately causes clouds to fragment in our simulations with isotropic conduction. In \citet{Kim2013}, the non-linear DLI saturates when the increased evaporation flow at convex regions equals the increased mass flux causing the advancing protrusion. In our simulations with more realistic clouds, the combination of evaporating convex regions and condensing concave regions can generate large enough pressure gradients between the proximal regions to cause large mass flux into the concavities. The configuration of these over and under pressured regions influences mass flow into the cusps. The influx of mass tends to pinch the cold cloud directly behind the protrusion creating a bulb until the protrusion separates from the larger cloud structure, as shown in figure \ref{fig:dli}. Once the protrusion separates, its size and shape determines whether it completely evaporates or survives long enough to merge with a larger cloud. In this way, the DLI in the turbulent 2D environment tends to continually disrupt large clouds to form smaller clouds. 

Although the motions in our simulations are ultimately driven by evaporation and condensation, viscosity also plays a role in shaping cloud dynamics. \citet{Yatou2009} performed multidimensional calculations and 1D simulations to show the increased survivability of viscous clouds. They suggest that the balance of viscosity with pressure gradients across transition layers greatly suppresses the evaporation of cold clouds. We see this in our simulations as well. Distortions formed on the cloud surfaces through DLI occur on smaller scales with decreasing Prandlt number. In this way viscosity acts like surface tension on the clouds. Once cloudlets break off, survival times strongly depend on Prandlt number. In the $\text{Pr} = 10^{-2}$ run, ejected cloudlets are so small and the evaporation rates are so high that they dissipate rapidly. In the $\text{Pr} = 2/3$ run, cloudlets ejected from larger clouds can survive long enough to merge into another large cloud (Figure~\ref{fig:iso}).

\section{Magnetohydrodynamic Simulations} \label{sec:mhd}
The addition of magnetic fields further expands the parameter space. Before focusing on the effects of different magnetic field strengths, we first do a preliminary investigation at fixed initial plasma beta $\beta = P/P_{\text{mag}}$ with varying Prandtl numbers. We find that there are marginal variations in cloud structures with different Prandtl numbers but with a noticeable difference in the evolution rate. Higher Prandtl number runs evolve more slowly due to reduced velocities. We choose a fixed $\text{Pr} = 10^{-2}$ for all simulations discussed in the following of this section.

\subsection{The Effects of Magnetic Fields on Linear TI} \label{sec:linear_mhd}

\begin{figure}
    \centering
    \includegraphics[width=\columnwidth]{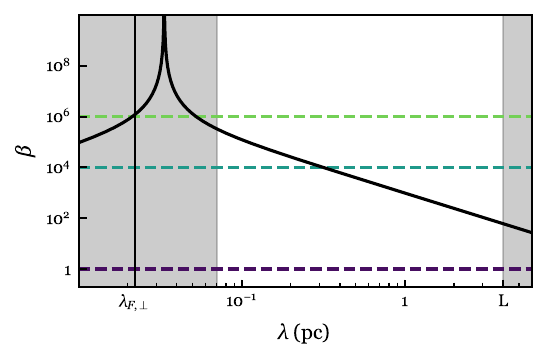}
    \caption{The solid black curve shows $\beta_{\text{crit}}$ as a function of perturbation wavelength $\lambda$. For a given $\lambda$, strong magnetic fields with $\beta \ll \beta_{\text{crit}}$ will suppress perturbations perpendicular to the fields. The vertical line indicates the Field length perpendicular to the magnetic field. The unshaded region shows the range of wavelengths initialized by our GRF perturbations. The dashed horizontal lines show the initial plasma betas chosen for our MHD runs: (i) for $\beta = 10^6$ no wavelengths will be suppressed and the initial linear phase of the TI will proceed similarly to the hydrodynamic case; (ii) for  $\beta = 10^4$ the smallest wavelengths will be suppressed and (iii) for the equipartition value $\beta =1$ all wavelengths will be completely suppressed.}
    \label{fig:beta_crit}
\end{figure}
The effect magnetic fields have on the linear development of TI was studied by \citet{Field1965}. He concluded that for sufficiently strong magnetic fields, collapse of condensation modes is suppressed for $\Vec{k}$ perpendicular to $\Vec{B}$ and largely unobstructed for the parallel case. We should expect motions to be predominately in the direction along the field lines when
\begin{equation} \label{ani suppressed}
    k_{\rho}-k_{T}' \ll \frac{v_a}{c_s}k\;,
\end{equation}
where the Alfv\'en velocity $v_a = B/\sqrt{4\pi\rho}$. The wavenumber $k_T$ is now modified by perpendicular and parallel thermal conduction coefficients,
\begin{equation} \label{kt'}
    k_{T}' = k_T + k^2\left(\frac{\cos^2{\theta}}{k_{\kappa_{\parallel}}}+\frac{\sin^2{\theta}}{k_{\kappa_{\bot}}} \right)\;,
\end{equation}
where $\theta$ is the angle between $\Vec{k}$ and $\Vec{B}$. In the notation previously introduced and in the limit $\kappa_{\text{iso}} \ll \kappa_{\text{aniso}}$ we can approximate $k_{\kappa_{\bot}} \approx \kappa_{\text{iso}}$ and $k_{\kappa_{\parallel}} \approx \kappa_{\text{aniso}}$. 

\begin{figure*}
    \centering
    \includegraphics[width=\textwidth]{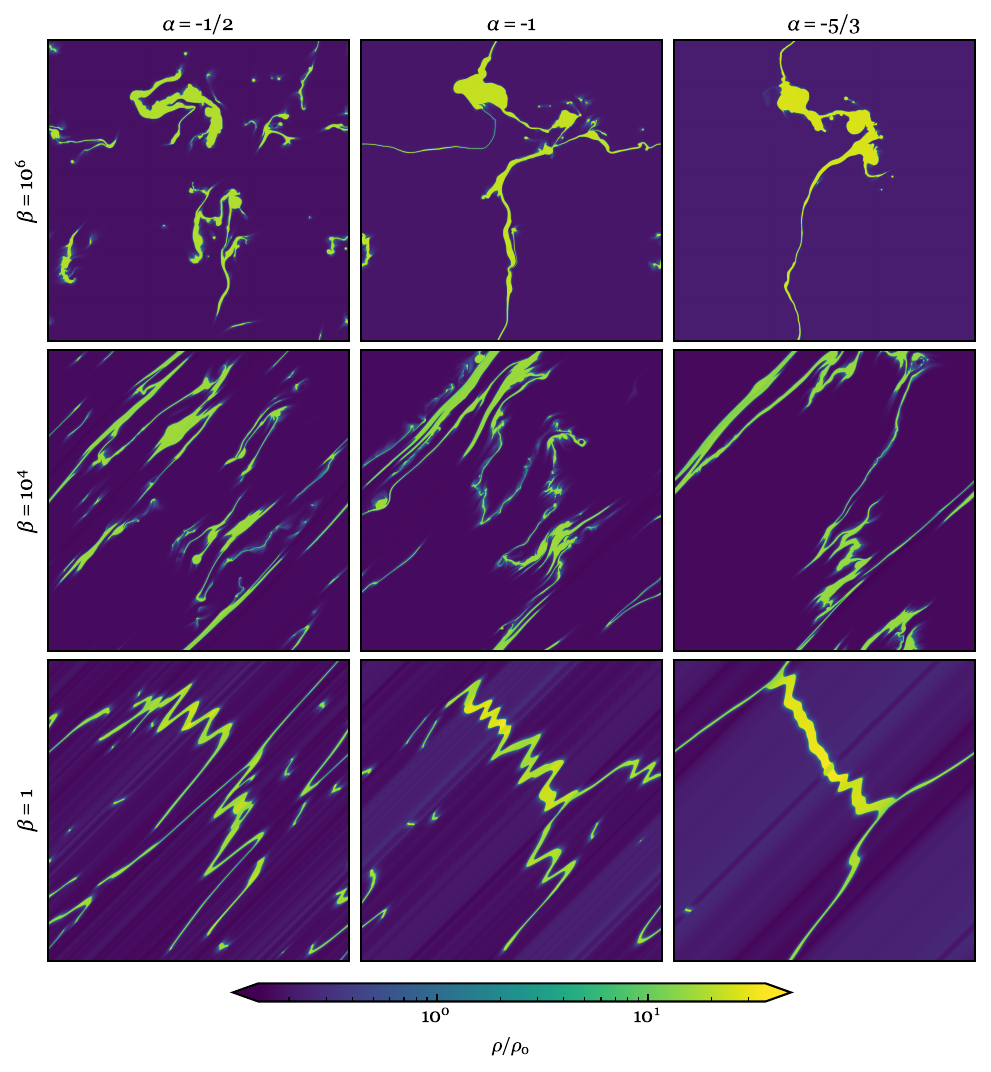}
    \caption{Late-time ($t/t_{\text{cool}} \approx 100$) density snapshots $\rho/\rho_0$ of MHD runs with varying spectral index $\alpha$ and plasma beta $\beta$. All runs produce dense clouds with filamentary structures, in contrast to the non-MHD runs. In the non-linear phase, the cold structures are determined by both the initial $\beta$ and $\alpha$.}
    \label{fig:beta_alpha}
\end{figure*}

\begin{figure*}
    \centering
    \includegraphics[width=\textwidth]{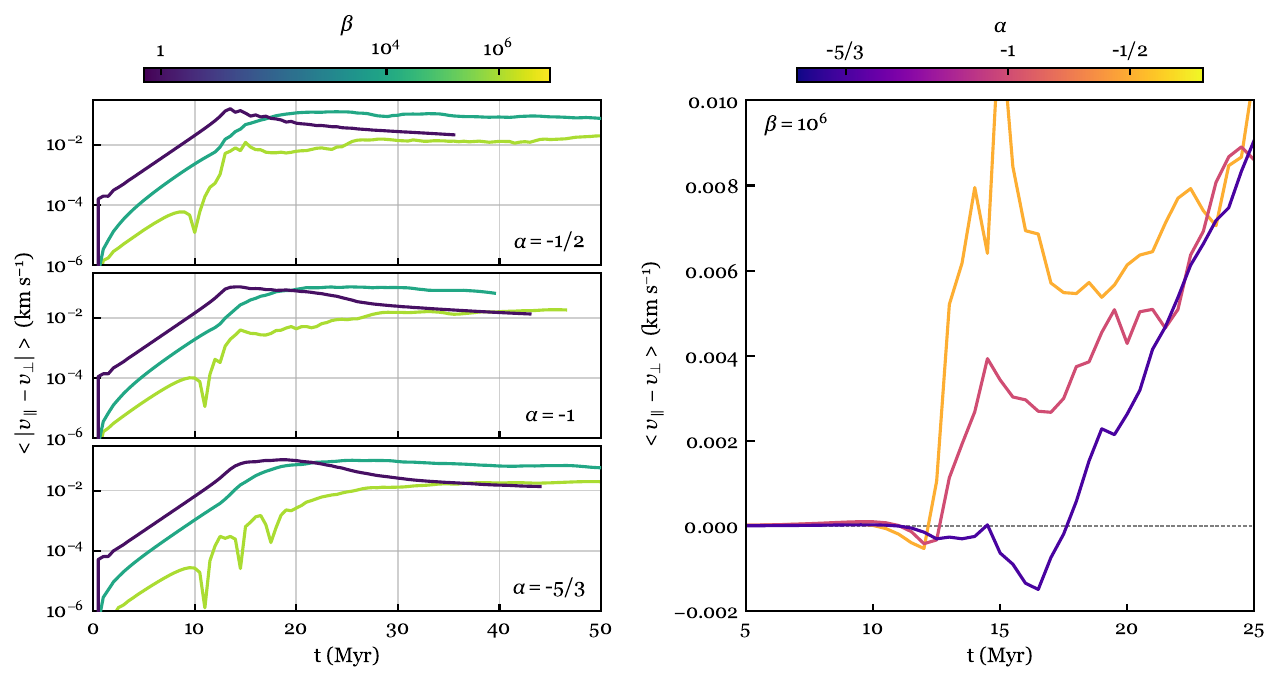}
    \caption{Left: time evolution of the mean absolute value of velocity component differences. 
    At early times, the linear cross-field suppression causes the largest velocity component difference at $\beta = 1$ and the smallest difference at $\beta = 10^6$. The $\beta = 10^6$ runs exhibit oscillatory behavior with multiple ``pulses''. Right: time evolution of the mean velocity component differences of the $\beta = 10^6$ run for each spectral index. Strong early oscillations result in multiple zero crossings for the $\alpha = -5/3$ run.}
    \label{fig:vmag}
\end{figure*}

For condensations across field lines, $\cos{\theta} \neq 1$, magnetic fields are compressed and amplified. Using equation (\ref{ani suppressed}) with our initial conditions we can estimate which choices of initial $\beta$ will cause significant TI suppression perpendicular to the magnetic field. For $\theta = \frac{\pi}{2}$ we derive 
\begin{equation} \label{beta inf}
    \beta_{\text{crit}}(k) = \frac{2}{\gamma}\left(\frac{k_{\rho}}{k}-\frac{k_T}{k}-\frac{k}{\kappa_{\bot}} \right)^{-2}\;,
\end{equation}
such that all $\beta(k) \ll \beta_{\text{crit}}(k)$ with wavenumber $k$ perturbations are suppressed. This critical beta as a function of wavelength is shown in Figure \ref{fig:beta_crit}. We choose initial plasma betas to represent both extreme regimes of no perpendicular suppression ($\beta = 10^6$) and complete perpendicular suppression ($\beta = 1$). We also run a simulation at an intermediate plasma beta ($\beta = 10^4$) that has small wavelength suppression. Interestingly, there exists a wavenumber,
\begin{equation} \label{k inf}
    k_{\infty} = \sqrt{\frac{\rho^2}{\kappa_{\bot}}\left(\frac{\mathcal{L}_{\rho}}{T} - \frac{\mathcal{L}_T}{\rho} \right)}\;,
\end{equation}
where any choice of $\beta$ leads to complete cross field TI suppression. This perturbation wavelength does exist above the perpendicular Field length $\lambda_{F,\bot}$. However, our choice of wavenumber cutoff $k_{\text{max}}$ is such that we do not seed the TI with wavenumber perturbations of $k_{\infty}$. Therefore, it is not explored further in this paper.

\subsection{MHD Simulations with Anisotropic Conduction and Viscosity}

We now present the nonlinear late-time results of MHD TI. Figure \ref{fig:beta_alpha} shows density ($\rho/\rho_0$) snapshots for varying spectral index and plasma beta. Dense clouds formed from TI show filamentary structures for all the runs but with the addition of some clumpy structures for the $\beta = 10^6$ runs. The structures of the $\beta =1$ run overall are more orderly and smooth and display less diversity, whereas the larger $\beta$ runs show a variety of globular, looping, and jagged structures. Noticeably, magnetic fields drastically alter the cold structures in the non-linear phase compared to the isotropic simulations even when the magnetic fields are weak ($\beta=10^4$ and $\beta=10^6$).

\subsubsection{Suppression of Cross-field Motion in TI with MHD}

Figure \ref{fig:vmag} shows the time evolution of the mean velocity component difference $\left<|v_{\parallel}-v_{\bot}|\right>$, and confirms the linear cross-field suppression prediction discussed in Section \ref{sec:linear_mhd}. The linear analysis from $\beta_{\text{crit}}(k)$ shown in Figure \ref{fig:beta_crit} predicts that the cross-field velocity $v_{\bot}$ should decrease with decreasing $\beta$ as TI contraction is suppressed. The difference between velocity components is the largest for $\beta =1$ jumping to $\sim 10^{-4}$ km s$^{-1}$ right after the first timestep, suggesting that the linear evolution is dominated by motions along the field lines. The velocity component difference at early times ($t<15$ Myr) decreases as $\beta$ increases. Interestingly, for only the $\beta = 10^6$ runs, oscillatory behavior is seen with 'pulses'. Rewriting equation~(\ref{pressure_ani}) in terms of the plasma beta we see that our simulations are susceptible to the firehose instability when $P_{\parallel} \gtrsim P_{\bot} + 2P/\beta$, which is most easily satisfied by our $\beta = 10^6$ runs. When a kink forms in a magnetic field line, the excess parallel pressure will accentuate and grow the kink causing an increase in perpendicular velocity at the instability location. The rapid growth of the instability will increase the perpendicular pressure, effectively regulating the pressure anisotropy. This is most likely the cause of the 'pulses'. 

All simulations saturate as velocities begin to be dominated by thermal conduction. In all spectral index groups the $\beta =1$ saturates first. The shape around the saturation point tends to smooth and become more gradual with decreasing $|\alpha|$ due to a slower evolution for isobaric condensations. As discussed in Section \ref{sec:nocond}, the larger wavelengths, $\lambda > \lambda_{\text{cool}}$, in the larger $|\alpha|$ runs are able to evolve isochorically where pressure gradients persist and drive larger velocities.

\subsubsection{Alignment between Density Structures and Magnetic Fields}

We calculate the anisotropy of the density fields to quantify the correlation between filament orientation and field direction. We adopt the parallel and perpendicular length scales of the density field introduced by \citet{Sharma2010},
\begin{equation} \label{l par}
    L_{\parallel} \equiv \frac{\int |\delta \rho| dV}{\int \left|\bm{\hat{b}} \cdot \bm{\nabla} \delta \rho \right| dV}\;,
\end{equation}
\begin{equation} \label{l per}
    L_{\bot} \equiv \frac{\int |\delta \rho| dV}{\int \left|(\bm{\hat{z}} \times \bm{\hat{b}}) \cdot \bm{\nabla} \delta \rho \right| dV}\;,
\end{equation}
where $\delta \rho = \rho - \left<\rho \right>$, $\left<\rho \right>$ is the volume averaged density, and $\bm{\hat{z}}$ is the unit vector pointing perpendicularly out of the computational domain. For $L_{\parallel}/L_{\bot} > 1$ the cold filaments are preferentially aligned with the magnetic fields. For $L_{\parallel}/L_{\bot} < 1$ the cold filaments are preferentially aligned across the magnetic fields. 

In the saturation state, all our MHD runs have $L_{\parallel}/L_{\bot}$ falling between $\sim 1.5 - 15$, suggesting some level of alignments. The highest $L_{\parallel}/L_{\bot}$ is seen in the $\beta=10^6$, $\alpha = -5/3$ run. The $\beta = 1$ runs all produce $L_{\parallel}/L_{\bot}$ just around 1. $L_{\parallel}/L_{\bot} \approx 1$ can have two interpretations: (1) density structures are agnostic to the magnetic field directions, or (2) density structures exist in two populations of parallel and perpendicular alignment to the magnetic field with $L_{\parallel} \approx L_{\bot}$. The second case is the cause of the $\beta =1$ anisotropy hovering around unity. For example, the bottom right panel of Figure~\ref{fig:beta_alpha} shows two thin filaments aligned with the magnetic fields and one thick filament perpendicular to the field direction.

Density features have also been shown to align with magnetic fields in previous turbulence simulations, and have been used observationally to infer the magnetic field orientation both in the ISM and in galaxy clusters \citep{Soler2013, Alex2017, Hu2020}.

\subsubsection{Evolution of Magnetic Field Orientation}

\begin{figure}
    \centering
    \includegraphics[width=\columnwidth]{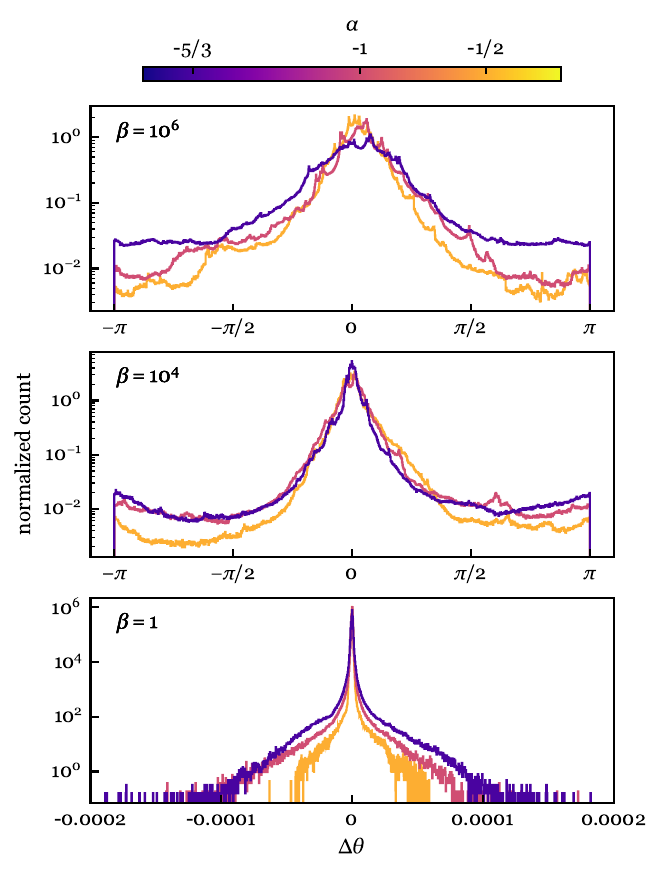}
    \caption{Histograms of the change in field direction at late-time snapshots. Each panel shows runs with different $\beta$. Within each panel the different spectral indices $\alpha$ are compared. The distributions peaks at zero, and narrow with increasing $\beta$. Strong magnetic tension along with complete cross-field TI suppression is enough to maintain the field's original orientation of $\theta_0 = 45^{\circ}$ in the $\beta =1$ run. Larger $\beta$ produces wider distributions of $\Delta \theta$, with magnetic field orientations completely reversed at $|\Delta \theta| = \pi$ in parts of the simulation domain. For fixed $\beta$, increasing $|\alpha|$ results in more field direction change. }
    \label{fig:delta_theta}
\end{figure}

The evolution of TI can result in a change in the magnetic field orientation $\theta$. In Figure \ref{fig:delta_theta}, we show the distributions of $\Delta \theta \equiv \theta - \theta_0$ for late-time snapshots of our MHD runs with different $\alpha$ and $\beta$. All runs have distributions that peak at $\Delta \theta = 0$ with varying widths. The $\beta = 1$ runs have extremely sharp peaked distributions with wing widths $|\Delta \theta| < 0.0002$, showing that the original field orientation is robustly preserved through TI collapse and nonlinear evolution. Flux-freezing causes all gas motions to remain within magnetic field ``lanes'' oriented at $\theta_0$. Striations in both the CNM and WNM are seen parallel to the field as gas cannot condense or evaporate across neighboring lanes due to anisotropic heat transport. In this way a majority of filament structures are oriented along or at right angles to the initial field. 

The distribution broadens with increasing $\beta$ as weaker magnetic tension and less cross-field TI suppression allows for significant field redirection. The $\beta = 10^4,10^6$ runs have significantly wider distributions with magnetic field orientations spanning $0 \leq |\Delta \theta| \leq \pi$. This allows more diverse cloud shapes for higher $\beta$ runs compared to the equipartition $\beta$ run, as is shown in Figure~\ref{fig:beta_alpha}. Each $\beta$ shows a subtle trend of greater field direction change for increasing $|\alpha|$. As discussed in Section~\ref{sec:nocond}, larger thermal pressure gradients in isochoric collapsing are able to shift the field direction more than isobaric cases.

\section{Discussions \& Conclusions} \label{sec:conclusion}

In this work we investigate TI in the ISM by conducting 2D hydrodynamic and MHD simulations spanning a wide parameter space. We first examine the effect of initial perturbations on the evolution through both the linear and non-linear regime of TI. We analyze cloud fragmentation in relation to recent work including the ``shattering'' and ``splattering'' views, and identify the mechanisms of cloud disruption. We study the effects of magnetic fields on the motions and the alignments of the cold clouds, and how TI affects field orientations. We summarize our findings as follows.
\begin{enumerate}
  \item In pure hydrodynamic runs without thermal conduction or physical viscosity, we vary the initial power spectral index $\alpha$ of the GRF density perturbations and find dramatic difference in the size, shape, dynamics and number of cloud structures formed in the simulations. 
  For our highest values of $|\alpha|$ one large cloud forms out of pressure balance due to the growth of unstable wavelengths $\lambda > \lambda_{\text{cool}}$.   
  For low values of $|\alpha|$ where power is concentrated in wavelengths $\lambda < \lambda_{\text{cool}}$, a large number of small, isobaric clouds form throughout the domain with slow translational motion. However, inevitable cloud merging at late times shows a tendency towards similar one cloud configuration across all sprectral indices, in agreement with previous works.
  
  \item When isotropic thermal conduction and physical viscosity are included, these processes dominate the evolution of the system and the initial $\alpha$ plays little role in the cloud structure in the non-linear phase of TI. The motions are mainly driven by the evaporation flows at the cold-warm interfacial layers, and the speed is affected by viscosity. Overall, conduction determines the final outcome of the TI, while viscosity largely controls the rate of the evolution. 
  
  \item In pure hydrodynamic runs without thermal conduction and physical viscosity, cloud disruption can occur as a result of vorticity generated via two mechanisms: the first is akin to the Richtmyer--Meshkov instability when a protrusion of cold gas is launched through the warm expanding medium after contraction rebound; the other is a tearing process as a by-product of opposing coalescent influences between two clouds.
  
  \item For the runs including isotropic thermal conduction and physical viscosity, cloud disruption is a result of the DLI. 
  Cloud surfaces experience stimulated evaporation or condensation based on the curvature. When a finger-like protrusion is formed via the DLI, the geometric term of the thermal conduction drives cloud motions that tend to pinch off clouds at the ends.
  
  \item We carry out MHD runs with anisotropic thermal conduction and physical viscosity. The initial magnetic fields are uniform with a range of plasma $\beta$ from 1 to $10^6$. Low $\beta$ suppresses collapse across field lines in the linear phase of TI, and gas motions are trapped within ``lanes'' along the field due to flux freezing. Although high $\beta$ runs with $\beta = 10^4$ and $10^6$ are still drastically different from the non-MHD runs in the non-linear phase. The combination of initial $\beta$ and GRF spectral index $\alpha$ determines the resultant cloud structures and filament orientations. We derive a critical value $\beta_{\text{crit}}$ to demarcate when complete linear TI suppression occurs.
  
  \item All our MHD runs show some level of alignment between dense filaments and the local magnetic fields. In the equipartition case ($\beta=1$), filaments are either parallel or perpendicular to the magnetic fields. The evolution of TI can also reorient magnetic fields. In high $\beta$ runs, significant field reorientation occurs.

\end{enumerate}

 Tiny \ion{H}{i} clouds have been observed to have sizes $\sim 0.01$ pc \citep{Braun2005,Stanimirovic2005}. Most of our simulations discussed here produce cold clouds with features of various sizes. We have discussed several mechanisms for cloud fragmentation in different physical conditions, which may be possible explanations for the observed tiny cold clouds. Because cloudlet production through DLI feeds off of geometric disturbances on the cloud surfaces, fragmentation is expected to be more efficient in realistic environments with stellar feedback driven turbulence.

We see in our MHD runs (Figure~\ref{fig:beta_alpha}) populations of both parallel and perpendicular alignment between dense filaments and local magnetic fields, which have also been shown to exist in observations. For example, in the Taurus molecular cloud \citep{Goldsmith2008}, the DR21 ridge in Cygnus X \citep{Schneider2010,Hennemann2012} and the Musca cloud \citep{Cox2016} main dense filaments are found to be oriented perpendicularly to the local field direction while lower density striations are found parallel to the field. Although direct extrapolations from our idealized 2D simulations to real ISM conditions should be made conservatively, our $\beta = 1 \ll \beta_{\text{crit}}$ simulations produce structures suggestive of the filament systems in Taurus, DR21 and Musca. The low density striations in our simulations act as lanes along which material flows onto the main filament, supplying it with new mass. The continuous influx of fresh mass allows these filaments to be active sights of new star formation in the real ISM. Theoretical arguments \citep{Nagai1998} and numerical simulations \citep{Nakamura2008} have shown that this filament configuration is favorable for self gravitating cold gas threaded with a constant magnetic field. Our results suggest that this outcome may be produced through TI in the presence of anisotropic conduction, even in the absence of gravity. These two processes (TI and gravity) most likely work together to form the observed filament structures observed in the ISM.

Several limitations to our work should be addressed in future investigations. First, the choice of constant diffusion coefficients instead of the physically motivated temperature dependent coefficients alters the evaporation/condensation front structure between the CNM and WNM. Our constant thermal conduction coefficient exaggerates the width of the transition layer and the mass flux across the layer. A more detailed study on the effects of temperature dependant conduction and viscosity coefficients on the DLI and subsequent cloud production should be an area of interest for future studies.

Second, the constant ratio of anisotropic to isotropic diffusion coefficients is chosen for simplicity and does not reflect the true microphysics of a magnetized plasma. As the CNM forms from the TI, gas temperatures drop causing the gyroradius and mean-free-path of particles to become comparable and thus perpendicular collisions become more frequent. The degree of plasma anisotropy should be temperature dependent. The possible effects this may have on cloud structures and dynamics should be investigated in future studies.

Finally, important cloud forming processes identified in this paper may be exaggerated in 2D. Due to a lack of a third dimension, 2D vorticity magnitude is accentuated and plays in favor of cloud shattering. In 3D, cloud breakup due to strong vorticities may be less frequent. For 2D DLI, the lack of a third dimension again plays in favor of cloud disruption. Because the growth of the DLI is a balancing act between the surface area of the finger and the mass flux across the front, dimensionality should play a significant role in the instability evolution. A finger's surface area grows much quicker in 3D than in 2D and thus the DLI may have difficulty producing smaller clouds in the manner discussed in this paper. Although the simplicity of a 2D system allows for ease of interpretation and insight into the physics at play, it can be hard to predict other results, especially in the MHD case, in 3D. Therefore, future investigations should explore the 3D case.

\section*{Acknowledgements}
We acknowledge helpful discussions with Eliot Quataert, Chris McKee, Greg Bryan, Jim Stone, Drummond Fielding and Yan-fei Jiang. We would like to thank Philipp Kempski for sharing his anisotropic viscosity module, and Ena Choi, Sean Ressler, Kareem El-Badry, and Andrea Antoni for providing support during the start of this project. The simulations were run on the NASA Pleiades supercomputer through allocation HEC-SMD-19-2293 and on the Berkeley Research Computing Savio cluster at the University of California, Berkeley. YL acknowledges support from NASA through Chandra Theory Grant TM8-19009X. This research made use of the open source project {\tt yt}\footnote{\href{https://yt-project.org}{yt-project.org}} \citep{Turk2011}.

\section*{Data Availability}
Animations of our simulations are available at \href{https://michaeljennings11.github.io}{https://michaeljennings11.github.io}. Data supporting our findings are available upon request to the corresponding author.

%%%%%%%%%%%%%%%%%%%%%%%%%%%%%%%%%%%%%%%%%%%%%%%%%%



\bibliographystyle{mnras}
\bibliography{TI} % if your bibtex file is called example.bib


%%%%%%%%%%%%%%%%%%%%%%%%%%%%%%%%%%%%%%%%%%%%%%%%%%

%%%%%%%%%%%%%%%%% APPENDICES %%%%%%%%%%%%%%%%%%%%%

%%%%%%%%%%%%%%%%%%%%%%%%%%%%%%%%%%%%%%%%%%%%%%%%%%

% Don't change these lines
\bsp	% typesetting comment
\label{lastpage}
\end{document}